\documentclass[3p,times,procedia]{elsarticle}
\usepackage{nupha_ecrc}

\volume{00}

\firstpage{1}

\journalname{Nuclear Physics A}

\runauth{}

\jid{nupha}

\jnltitlelogo{Nuclear Physics A}

\usepackage{amssymb}

\usepackage[figuresright]{rotating}

\usepackage{color}
\usepackage{braket}
\usepackage[mathscr]{eucal}

\newcommand{\beq}{\begin{equation}}
\newcommand{\eeq}{\end{equation}}

\newcommand{\n}{\nonumber}
\newcommand{\x}{x_1}
\newcommand{\y}{x_2}
\newcommand{\s}{y}
\newcommand{\X}{x}
\newcommand{\F}{\mathcal{F}}
\newcommand{\Pc}{\mathcal{P}}
\newcommand{\V}{\mathcal{V}}
\newcommand{\A}{\mathcal{A}}
\newcommand{\Sc}{\mathcal{S}}

\begin{document}

\begin{frontmatter}



\dochead{XXVIIIth International Conference on Ultrarelativistic Nucleus-Nucleus Collisions\\ (Quark Matter 2019)}

\title{Wigner function and kinetic theory for massive spin-1/2 particles}


\author[GU]{Nora Weickgenannt}
\author[GU,HU]{Xin-li Sheng}
\author[GU]{Enrico Speranza}
\author[HU]{Qun Wang}
\author[GU,HU]{\, \, \, \, \, \, \, \, \, Dirk H. Rischke}
\address[GU]{Institute  for  Theoretical  Physics,  Goethe  University,
Max-von-Laue-Str.\  1,  D-60438  Frankfurt  am  Main,  Germany}
\address[HU]{Interdisciplinary Center for Theoretical Study and Department of
Modern Physics, University of Science and Technology of China, Hefei,
Anhui 230026, China}

\begin{abstract}
We calculate the Wigner function for
massive spin-1/2 particles in an inhomogeneous electromagnetic field to leading order in the Planck constant $\hbar$. 
Going beyond leading order in $\hbar$ we then derive a generalized Boltzmann equation 
in which the force 
exerted by an inhomogeneous electromagnetic field on the particle dipole moment 
arises naturally. Furthermore, a kinetic equation for this dipole moment is derived. Carefully taking the massless
limit we find agreement with previous results.
The case of global equilibrium with rotation is also studied. 
Our framework can be used to study polarization effects induced by vorticity and magnetic field in 
relativistic heavy-ion collisions.
\end{abstract}


\end{frontmatter}


\section*{}
In the recent years, there has been an intense theoretical activity which has led to a deeper understanding of 
{the} transport properties of
chiral matter, see e.g. Ref.\, \cite{Kharzeev:2015znc} for a review. However, only few attempts have been made to derive 
a covariant kinetic {theory} for \textit{massive} particles
using Wigner functions~\cite{Fang:2016vpj,Florkowski:2018ahw}. The aim of our work is to 
\textcolor{black}{fill this gap. Here we report on a recent paper~\cite{Weickgenannt:2019dks} where the} kinetic theory
for massive spin-1/2 particles in an inhomogeneous electromagnetic field was derived. This framework provides a basis
to study polarization effects in relativistic heavy-ion collisions \cite{STAR:2017ckg}. Our starting point is the covariant Wigner
function~\cite{Heinz:1983nx,Vasak:1987um,Wang:2001dm}. In order to solve the equations of motion for the Wigner function,
we employ an expansion in the Planck constant $\hbar$ and truncate at the lowest non-trivial order.

\section{Wigner function for massive spin-1/2 particles}
\label{EW}

The Wigner function is defined as the Fourier transform of the two-point correlation function \cite{Elze:1986qd},
\begin{eqnarray}
W_{\alpha\beta}(\X,p)& =& \int \frac{d^4 \s}{(2\pi\hbar)^4} e^{-\frac{i}{\hbar}p\cdot \s}   \left\langle :\bar{\psi}_\beta(\x)U(\x,\y)\psi_\alpha (\y): \right\rangle\,. \label{Wignerdef}
\end{eqnarray}
Here, $\x$ and $\y$ are the space-time coordinates of two different points, with $\s^\mu\equiv \x^\mu-\y^\mu$ and
$\X^\mu\equiv (\x^\mu+\y^\mu)/2$ and $U(x_1,x_2)$ is a gauge link.
In this paper, the electromagnetic field $\mathbb{A}_\mu$ will be treated as an external, classical field. \color{black} Under this assumption \color{black} one can derive the \textcolor{black}{exact} kinetic equation for the Wigner function~\cite{Elze:1986qd}:
\begin{equation}
(\gamma \cdot K - m)W(x,p)=0\,. \label{Dirac}
\end{equation}
Here one has defined the operator
$
K^\mu\equiv \Pi^\mu+\frac12i\hbar\nabla^\mu\,,$
with the generalized space-time derivative and momentum operators
$
\nabla^\mu\equiv\partial_x^\mu-j_0(\Delta)F^{\mu\nu}\partial_{p\nu}\, $ and 
$\Pi^\mu\equiv p^\mu-\frac{\hbar}{2} j_1(\Delta)F^{\mu\nu}\partial_{p\nu}\,,$ 
where $\Delta \equiv \frac{\hbar}{2}\, \partial_p\cdot\partial_x$ and $F^{\mu \nu} = \partial^\mu_x \mathbb{A}^\nu
- \partial^\nu_x \mathbb{A}^\mu$ is the electromagnetic
field-strength tensor. We should emphasize that in Eq.\ (\ref{Dirac}) the space-time derivative
$\partial_x$ contained in $\Delta$ only acts on  $F^{\mu\nu}$, but not on the Wigner function.
The functions $j_0(x)=\sin x /x$ and $j_1(x)=(\sin x-x \cos x) /x^2$ are spherical Bessel functions.

In order to derive a kinetic equation for massive spin-1/2 particles, it is advantageous to decompose the Wigner function
in terms of a basis formed by the 16 independent generators of the Clifford algebra,
\begin{equation}
W=\frac14\left(\F+i\gamma^5\Pc+\gamma^\mu \V_\mu+\gamma^5\gamma^\mu \A_\mu
+\frac12\sigma^{\mu\nu}\Sc_{\mu\nu}\right)\,. \label{dec}
\end{equation}
The coefficients $\F, \Pc, \V^\mu, \A^\mu,$ and $\Sc^{\mu\nu}$ correspond to the scalar, pseudo-scalar, vector, axial-vector, and tensor part of the Wigner function, respectively. They will be determined by solving Eq.\ (\ref{Dirac})  employing an expansion in $\hbar$.
 

\section{General solution and kinetic equations up to order $\hbar$}
\color{black} The components of the Wigner function are not all independent. We choose to express $\Pc$, $\V^\mu$, and $\A^\mu$ in terms of $\F$ and $\Sc^{\mu\nu}$.
\textcolor{black}{From the equations for the Wigner-function components we} \color{black} obtain the modified on-shell conditions for the scalar and
tensor components,
\begin{eqnarray}
(\Pi \cdot \Pi-m^2)\F&=&\frac{\hbar}{2}\Pi^\mu\nabla^\nu \Sc_{\nu\mu}\,, \n\\
(\Pi \cdot \Pi-m^2)\Sc_{\mu\nu}&=&-\Pi^\alpha \Pi_{[\mu}\Sc_{\nu]\alpha}
-\frac{\hbar}{2}\nabla_{[\mu}\Pi_{\nu]}\F -\frac{\hbar^2}{4}\nabla_{[\mu} \nabla^\alpha  \Sc_{\nu]\alpha}
+ \frac{\hbar}{2}\epsilon_{\mu\nu\alpha\beta}\Pi^\alpha \nabla^\beta\Pc\,,\label{on-shell}
\end{eqnarray}
\textcolor{black}{
where $A_{[\mu}B_{\nu]}\equiv A_\mu B_\nu-A_\nu B_\mu$.
For the zeroth order we use the solution from Ref.\, \cite{DeGroot:1980dk}. The first-order solution is obtained by finding general solutions of the on-shell conditions and plugging the zeroth-order solution into the first-order equations.
Eventually the Wigner function to first order in $\hbar$
can be written as } \color{black}
\begin{eqnarray}
\mathcal{F} & = & m\left[V\,\delta(p^{2}-m^{2})-\frac{\hbar}{2}F^{\mu\nu}\bar{\Sigma}_{\mu\nu}\,\delta^{\prime}(p^{2}-m^{2})\right]+\mathcal{O}(\hbar^{2})\, ,\nonumber \\
\mathcal{P} & = & \frac{\hbar}{4m}\epsilon^{\mu\nu\alpha\beta}\nabla_{\mu}^{(0)}\left[p_{\nu}\bar{\Sigma}_{\alpha\beta}\,\delta(p^{2}-m^{2})\right]+\mathcal{O}(\hbar^{2})\, ,\nonumber \\
\mathcal{V}_{\mu} & = & p_{\mu}\left[V\,\delta(p^{2}-m^{2})-\frac{\hbar}{2}F^{\alpha\beta}\bar{\Sigma}_{\alpha\beta}\,\delta^{\prime}(p^{2}-m^{2})\right]+\frac{\hbar}{2}\nabla^{(0)\nu}\left[\bar{\Sigma}_{\mu\nu}\,\delta(p^{2}-m^{2})\right]+\mathcal{O}(\hbar^{2})\, ,\nonumber \\
\mathcal{A}_{\mu} & = & -\frac{1}{2}\epsilon_{\mu\nu\alpha\beta}p^{\nu}\left[\bar{\Sigma}^{\alpha\beta}\,\delta(p^{2}-m^{2})\right.-\left.\hbar F^{\alpha\beta}V\,\delta^{\prime}(p^{2}-m^{2})\right]+\mathcal{O}(\hbar^{2})\, ,\nonumber \\
\mathcal{S}_{\mu\nu} & = & m\left[\bar{\Sigma}_{\mu\nu}\,\delta(p^{2}-m^{2})-\hbar F_{\mu\nu}V\,\delta^{\prime}(p^{2}-m^{2})\right]+\mathcal{O}(\hbar^{2})\, .\label{eq:resummed-Wigner}
\end{eqnarray}
The undetermined functions
$V$ and $\bar{\Sigma}_{\mu\nu}$ satisfy one constraint equation,
\begin{equation} \label{eq:constraintt}
p^{\nu}\bar{\Sigma}_{\mu\nu}\, \delta(p^2-m^2)=\frac{\hbar}{2}\, \delta(p^2-m^2)\,
\nabla_{\mu}^{(0)}V+\mathcal{O}(\hbar^{2})\,,
\end{equation}
and two kinetic equations
\begin{eqnarray}
0 & = & \delta(p^{2}-m^{2})\left[ p\cdot\nabla^{(0)}{V}+\frac{\hbar}{4}(\partial_{x}^{\alpha}F^{\mu\nu})\partial_{p\alpha}{\bar{\Sigma}}_{\mu\nu}\right]+\mathcal{O}(\hbar^{2})\, ,\nonumber \\
0 & = & \delta(p^{2}-m^{2})\bigg[ p\cdot\nabla^{(0)}{\bar{\Sigma}}_{\mu\nu}-F_{\ [\mu}^{\alpha}{\bar{\Sigma}}_{\nu]\alpha} \bigg. \bigg. +\frac{\hbar}{2}(\partial_{x\alpha}F_{\mu\nu})\partial_{p}^{\alpha}{V} \bigg]+\mathcal{O}(\hbar^{2})\, . \label{eq:kin-after-transformation}
\end{eqnarray}
\color{black} From the results (\ref{eq:resummed-Wigner}) and (\ref{eq:kin-after-transformation}) \color{black}it is possible to derive fluid-dynamical equations of motion with spin degrees of freedom using the canonical definitions of \textcolor{black}{the} 
energy-momentum and spin tensors~\cite{Weickgenannt:2019dks}. In accordance with previous
works \cite{Florkowski:2017ruc,Becattini:2018duy}, the conservation of the total angular momentum is promoted as
an additional fluid-dynamical equation, where the divergence of the spin tensor is related to the antisymmetric part
of the energy-momentum tensor.

\section{Comparison to the massless and classical case}
\color{black} The on-shell dipole moment $\bar{\Sigma}_{\mu\nu}$ can be written at the zeroth order as $\bar{\Sigma}^{\mu\nu}=\Sigma^{\mu\nu}A$ where 
\begin{equation} \label{Sigma}
 \Sigma^{\mu\nu}=-\frac1m \epsilon^{\mu\nu\alpha\beta}p_\alpha n_\beta
\end{equation}
 is the dipole-moment tensor, $n_\beta$ is the polarization vector and $A$ is the spin-antisymmetric part of the distribution function. \color{black} On the classical level, $\Sigma^{\mu\nu}$ is the intrinsic angular-momentum tensor about the center
of mass. In a relativistic theory, the center of mass of a particle is frame-dependent. In order to have a
frame-independent definition of $\Sigma^{\mu \nu}$, one requires
$p_\mu \Sigma^{\mu\nu}=0$ as a gauge condition. This requirement identifies the dipole-moment tensor as the intrinsic angular-momentum tensor about the center of mass in the rest frame
of the particle \cite{Stone:2014fja}. For massless particles there is no rest frame, thus both the position (in the classical case the center of momentum)
and the dipole-moment tensor can at first be defined in an arbitrary frame characterized by a time-like four-vector $u^\mu$, which means that we choose the gauge condition
$u^\mu \Sigma_{\mu\nu}=0$~\cite{Chen:2015gta}.
Consequently, the frame vector $u^\mu$ must assume the role of $p^\mu$ in Eq.\ (\ref{Sigma}).
Moreover, since $n^\mu$ and $p^\mu$ are parallel for massless particles, the momentum $p^\mu$
can assume the role of $n^\mu$ in Eq.\ (\ref{Sigma}). Finally, in order to obtain the massless case
we need to replace the normalization factor $1/m$ in Eq.\ (\ref{Sigma}).
The energy of a massive on-shell particle in its rest frame is $p^0_{rf} = \sqrt{p^2}=m$.
The energy of a massless particle in the rest frame of $u^\mu$, however, is $p^0_u = p \cdot u$.
Thus, it is natural to replace the normalization $1/m$ in Eq.\ (\ref{Sigma}) by $1/(p\cdot u)$. We emphasize that this
replacement can only be done in the presence of a $\delta$-function which sets the rest-frame energy equal to the
mass $m$. \color{black} With these replacements we can show the agreement of our results for the vector and axial-vector current given in Eqs.\, (\ref{eq:resummed-Wigner}) in the massless limit with the previously known massless solution \cite{Hidaka:2016yjf,Huang:2018wdl,Gao:2018wmr}.
\color{black}

We show that Eq.\ (\ref{eq:kin-after-transformation}) gives rise to the first and second Mathisson--Papapetrou--Dixon (MPD)
equations  \cite{Bailey:1975fe,Israel:1978up} as well as to the Bargmann-Michel-Telegdi (BMT) equation  \cite{Bargmann:1959gz}, which
were derived for classical, extended, spinning particles with non-vanishing dipole moment.
Comparing Eq.\ (\ref{eq:kin-after-transformation}) to the generic form of the collisionless relativistic Boltzmann--Vlasov 
equation~\cite{Israel:1978up,Cercignani}
 we find that in our case the external force $F_s^\mu$ acting on a particle with spin up/down for $s=\pm$ is given as the sum of the Lorentz force and the Mathisson~force.
\begin{equation} \label{gen_force}
 F_s^\mu=\frac1m \left[F^{\mu\nu}p_\nu+s\frac\hbar4(\partial^\mu_x F^{\nu\rho})\Sigma_{\nu\rho}^{(0)}\right]\,,
\end{equation}
In Refs.\ \cite{Bailey:1975fe,Israel:1978up}, the first MPD equation for particles with classical dipole moment
$m^{\mu \nu}$ was derived. Our results agree with this, setting
$
 m_{\mu\nu}\longrightarrow g \mu_B \frac{s}{2}\Sigma_{\mu\nu}^{(0)}\,,
$
with Bohr's magneton $\mu_B \equiv \mathfrak{e} \hbar/(2 m)$, where $\mathfrak{e}$ is the electric charge, 
and the gyromagnetic ratio $g=2$, as expected
for Dirac particles with spin 1/2.

The evolution of the zeroth-order dipole-moment tensor is given by the second equation in (\ref{eq:kin-after-transformation}):
\begin{equation}
 m\dot{\Sigma}^{(0)}_{\mu\nu}\equiv m(\dot{x}^\alpha\partial_{x\alpha}+\dot{p}^\alpha\partial_{p\alpha})\Sigma^{(0)}_{\mu\nu}=F^{\alpha}_{\ [\mu}\Sigma^{(0)}_{\nu]\alpha}\,, \label{MPD2}
\end{equation}
with $F^\mu_{\textcolor{black}{s}}$ given by Eq.\ (\ref{gen_force}) to zeroth order. 
Equation (\ref{MPD2}) is identical to the second MPD equation \cite{Bailey:1975fe,Israel:1978up}.

\section{Global equilibrium}
In this section we will consider a special solution of Eqs.~(\ref{eq:kin-after-transformation}) obtained in global equilibrium with rigid rotation. Assuming the standard form of the
collision term, the distribution function in equilibrium must have the form \cite{Chen:2015gta,Israel:1978up}
$V^{eq}=\sum_s(e^{g_s}+1)^{-1}\,$,
with $g_s$ being a linear combination of the collisional invariants, namely, charge, kinetic momentum,
and total angular momentum.
\color{black}
Requiring that the Boltzmann equation (\ref{eq:kin-after-transformation}) is fulfilled, we obtain from the third equation (\ref{eq:resummed-Wigner})
\color{black}
\begin{eqnarray}
 \V^\mu&\!\!=&\!\!\frac{2}{(2\pi \hbar)^3}\sum_s\Bigg[\delta(p^2-m^2) \left(p^\mu-m s \frac\hbar2 \tilde{\omega}^{\mu\nu}
 n_\nu\frac{\partial}{\partial g_s^{(0)}}\right)\Bigg.\left.+ ms \hbar \tilde{F}^{\mu\nu}n_\nu\delta'(p^2-m^2)\right.\n\\
 &&\Bigg.-\frac{s\hbar}{2m}\delta(p^2-m^2) \epsilon^{\mu\nu\alpha\beta}p_{\alpha} 
 \left(\nabla^{(0)}_\nu n_\beta\right)\Bigg]\left[\theta(p^0)f_s^{(0)+}+\theta(-p^0)f_s^{(0)-}\right]+\mathcal{O}(\hbar^2)\,. \label{V_eq}
\end{eqnarray}
The term containing $\tilde{F}^{\mu\nu}=(1/2)\epsilon^{\mu\nu\alpha\beta}F_{\alpha\beta}$ in Eq.\ (\ref{V_eq}) is caused by off-shell effects and describes the
vector current induced by electromagnetic fields, which yields the analogue of the chiral magnetic effect in the case of
non-zero mass. On the other hand, the term containing the dual of the thermal vorticity $\tilde{\omega}^{\mu\nu}$ describes 
the current induced by vorticity and thus gives the analogue of the chiral vortical effect.

\color{black} The axial-vector current in global equilibrium can be obtained in a similar way. \color{black} One identifies three contributions to the axial-vector current in the 
massive case. One term describes the spin precession in the presence of an electromagnetic field according to
the BMT equation. The second term gives rise to the axial current in the direction of the vorticity, which is the analogue of the axial chiral vortical 
effect. Finally, the last term describes the axial current along the magnetic field, which is the analogue 
of the chiral separation effect. These terms are analogous to those found in Refs.\  
\cite{Fang:2016vpj,Becattini:2016gvu,Lin:2018aon}.

After completion of our work \cite{Weickgenannt:2019dks}, we became aware of related studies \cite{Gao:2019znl,Hattori:2019ahi}.

\ 

\textbf{Acknowledgements} \,  The work of D.H.R., X.-l.S., E.S., and N.W.\ is supported by the
Deutsche  For- \\ schungsgemeinschaft (DFG, German Research Foundation)
through the Collaborative Research Center CRC-TR 211 ``Strong-interaction matter
under extreme conditions'' -- project number 315477589 - TRR 211. D.H.R.\ acknowledges partial support
by the High-end Foreign Experts project GDW20167100136 of the State
Administration of Foreign Experts Affairs of China.
X.-l.S.\ is supported in part by China Scholarship Council. E.S.\ acknowledges support by BMBF
``Verbundprojekt: 05P2015 - ALICE at High Rate", and 
BMBF ``Forschungsprojekt: 05P2018 - Ausbau von ALICE am LHC (05P18RFCA1)".
Q.W.\ is supported in part by NSFC under Grant No.\ 11535012 and 11890713.





\bibliographystyle{elsarticle-num}
\bibliography{biblio_paper}{}






\end{document}